\documentclass[a4paper,11pt]{article}
\pdfoutput=1 

\usepackage{jheppub}                                         
\usepackage{color}
\usepackage{epstopdf}
\usepackage{rotating}
\usepackage{booktabs}
\usepackage{hvfloat}
\usepackage{float}
\usepackage{adjustbox}
\usepackage{graphicx}
\usepackage[left=4.5cm,right=1.5cm,top=5.5cm]{geometry}

\newcommand{\be}{\begin{equation}}
\newcommand{\ee}{\end{equation}}
\newcommand{\ba}{\begin{eqnarray}}
\newcommand{\ea}{\end{eqnarray}}

\newcommand{\beq}{\begin{equation}}
\newcommand{\eeq}{\end{equation}}
\newcommand{\beqa}{\begin{eqnarray}}
\newcommand{\eeqa}{\end{eqnarray}}

\title{\boldmath Thermodynamic Volume  of Cosmological Solitons}

\author[a]{Saoussen Mbarek}
\author[a]{and Robert B. Mann}

\affiliation[a]{Department of Physics and Astronomy, University of Waterloo, \\
Waterloo, Ontario, Canada, N2L 3G1, Canada}

\emailAdd{smbarek@uwaterloo.ca}
\emailAdd{rbmann@sciborg.uwaterloo.ca}

\abstract{We present explicit expressions of the thermodynamic volume inside and outside the cosmological horizon of Eguchi-Hanson solitons in general odd dimensions. These quantities are calculable and well-defined regardless of whether or not the regularity condition for the soliton is imposed. For the inner case, we show that the reverse isoperimetric inequality is not satisfied for general values of the soliton parameter $a$, though a narrow range exists for which the inequality does hold. For the outer case, we find that the mass $M_{out}$ satisfies the maximal mass conjecture and the volume is positive. We also show that, by requiring $M_{out}$ to yield the mass of  dS spacetime when the soliton parameter vanishes,  the associated cosmological volume is always positive.}

\begin{document} 
\maketitle
\flushbottom

\section{Introduction}

Understanding black holes as thermodynamic systems is a subject that continues to yield new insights
into gravitational physics, providing us with important clues as to the  nature of quantum gravity. Asymptotically Anti de-Sitter (AdS) black holes have been of particular interest in recent years, in part because of their  significance in
various proposed gauge-gravity dualities, but also because they have been exhibit thermodynamic behaviour analogous to
that in everyday life, a subject known as Black Hole Chemistry  \cite{Kubiznak:2014zwa}.

In Black Hole Chemistry the cosmological constant $\Lambda$ is regarded as a thermodynamic variable, extending the phase space of black hole thermodynamics  \cite{CreightonMann:1995}.  The mass of the black hole can be understood as enthalpy  \cite{Kastor2009}
and the cosmological constant as pressure, with a conjugate thermodynamic volume $V$ \cite{Caldarelli2000, Kastor2009, Dolan2010, Dolan2011a, Dolan2011, Dolan2012, Cvetic2010, Larranaga2011, Larranaga2012,Gibbons2012, Kubiznak2012, Gunasekaran2012, Belhaj2012,  Lu2012, Smailagic2012, Hendi2012}.   From this
perspective one can show that the celebrated Hawking-Page phase transition \cite{Hawking:1982dh}  can be understood
as being analogous to a solid/liquid phase transition  \cite{Kubiznak:2014zwa}, and more generally that 
the 4-dimensional Reissner-Nordstr\"om AdS black hole can be interpreted as a Van der Waals fluid
with the same critical exponents \cite{Kubiznak2012}.    Along with  more general Van der Waals behaviour with standard critical exponents  \cite{Dolan2013,Dolan2013a,Dolan2013b,Gunasekaran2012,Zou2013,Zou2014,Ma2013, Ma2014, Wei2014,Mo2014, Mo2014a, Mo2014b,Zhang2014,Liu2014, Liu2014a,Rajagopal:2014ewa,Delsate:2014zma,Hennigar:2015esa}, other `chemical' black hole behaviour was subsequently discovered, such as reentrant phase transitions \cite{Altamirano:2013ane}, tricritical points \cite{Altamirano:2013uqa}, Carnot cycles \cite{Johnson2014a}, isolated critical points \cite{Frassino2014,Dolan:2014vba}, extensions to black rings \cite{Altamirano:2014tva},
and superfluidity \cite{Hennigar:2016xwd}.

The role of the  thermodynamic volume $V$ is not yet fully understood.  It was originally conjectured to satisfy a relation known as the {\em Reverse Isoperimetric Inequality}~\cite{Cvetic:2010jb, Altamirano:2014tva}, which states that the isoperimetric ratio 
\be\label{eq:ipe-ratio}
\mathcal{R}=\left(\frac{(D-1) {V}}{\omega_{D-2}}\right)^{\frac{1}{D-1}}\left(\frac{\omega_{D-2}}{ {A}}\right)^{\frac{1}{D-2}}
\ee
always satisfies $\mathcal{R} \ge 1$, where ${A}$ is the horizon area, and $\omega_d$ stands for the area of 
the space orthogonal to constant $(t,r)$ surfaces. Physically it implies, for example,  that the black hole of given ``volume'' $V$ with maximal entropy is the Schwarzschild-AdS black hole. However a class of black holes has recently been found that violates this conjecture  \cite{Hennigar:2014cfa, Hennigar:2015wxa}, necessitating further investigation of the role and meaning of the volume \cite{Kubiznak:2015bya}.  The relationship of $V$ to other proposed notions of volume \cite{Armas:2015qsv,Couch:2016exn} is an ongoing subject of investigation.

Our knowledge of thermodynamic volume, and more generally the thermodynamic behaviour of asymptotically de Sitter (dS) black holes, for which $\Lambda>0$,  is  significantly more sparse \cite{Romans1992,MannRoss:1995,Li:2014ixn,Tian:2014ila,Azreg-Ainou:2014lua,Kubiznak:2015bya,Zhang:2016yek}. 
Yet their importance to cosmology and to a posited  duality between gravity in de Sitter space and conformal field theory \cite{Strominger:2001pn} make them important objects of investigation.  However this is a complex problem, since the absence of a Killing vector that is everywhere timelike outside the black hole horizon renders   a good notion of the asymptotic mass questionable. Furthermore, the presence of both a black hole horizon and a cosmological horizon yields two distinct temperatures, suggesting that the system is in a non-equilibrium state. This in turn leads to some ambiguity in interpreting the thermodynamic volume, since distinct volumes can be associated with each horizon.  In all known examples the reverse isoperimetric inequality $\mathcal{R} \ge 1$ holds separately for each; however if the volume is taken to be the naive geometric volume in between these horizons then the isoperimetric inequality holds
\cite{Dolan:2013ft}.
   
It would be preferable to study the `chemistry' of cosmological horizons in isolation.  For this we need a class of solutions that
are not of constant curvature and that have only a cosmological horizon.  Fortunately a broad class of such solutions exists:    Eguchi-Hanson de Sitter solitons \cite{Clarkson:2005qx}.   

 The Eguchi-Hanson (EH) metric  is a self-dual solution of the four-dimensional vacuum Euclidean Einstein equations \cite{Eguchi:1978xp}.  It has  odd-dimensional generalizations that were discovered few years ago \cite{Clarkson:2005qx} in Einstein gravity with a cosmological
constant. They are referred to as the Eguchi-Hanson solitons. 
For $\Lambda<0$ they are horizonless solutions that in five dimensions are asymptotic to $AdS_5/ Z_p$ $(p\geq3)$ and have Lorentzian signature, yielding a non-simply connected background manifold for the CFT boundary theory \cite{Clarkson:2006zk}.  Solutions in higher dimensions have a more complicated asymptotic geometry.   For $\Lambda> 0$ these solutions in any odd dimension have a single cosmological horizon, by which we mean that
they have a Killing vector $\partial/\partial t$ that becomes spacelike at sufficiently large distance from the origin.  Upon taking the 
mass to be the conserved quantity associated with this Killing vector at future infinity, and computing it using
the counterterm method  \cite{Ghezelbash:2001vs}, 
these solutions all satisfy a {\it maximal mass conjecture} \cite{Balasubramanian:2001nb}, whose implication is that  they all have mass less than that of pure de Sitter spacetime with the same asymptotics.

In this paper we  study the Eguchi-Hanson de Sitter (EHdS) solitons in the context of extended phase space thermodynamics. In this framework, we consider the cosmological constant as a thermodynamic variable equivalent to the pressure in the first law where
\begin{equation}\label{press}
P = - \frac{\Lambda}{8 \pi G}
\end{equation}
though for $\Lambda > 0$ this quantity is negative and so is perhaps best referred to as a tension.  Noting this, we shall continue
to refer to $P$ as pressure; its corresponding conjugate is the thermodynamic volume $V$ and is defined from geometric arguments \cite{Dolan:2013ft}.  It ensures the validity of the extended first law
\begin{equation}\label{firstgenD}
\delta M - T{dS}  - V \delta P = 0
\end{equation}
and (consistent with Eulerian scaling) renders the Smarr formula valid:
\begin{equation}\label{SmarrgenD}
(d-2) M- (d-1) T{S} + 2 VP = 0
\end{equation}
where the spacetime dimension is given by $(d+1)$.

Motivated by the above, we use the Eguchi-Hanson solitons in de Sitter space to investigate their 
thermodynamics and  cosmological volume in the context of extended phase space. The particular advantage afforded by these solutions is that, unlike the situation with de Sitter black holes,  thermodynamic equilibrium is satisfied.  
We find explicit expressions for the thermodynamic volume inside and outside the cosmological horizon. 
 For the inner case, the reverse isoperimetric inequality is   satisfied only for a small range of $a>\sqrt{3/4} \ell$ when a regularity condition for the soliton is not satisfied. For the outer case, an important role is played by a Casimir-like term that appears as an arbitrary constant in the first law and Smarr relation. 
We compare our results to those obtained using the counterterm method \cite{Clarkson:2005qx} 
and we find that they match. We note that for this case the mass is always smaller than maximal mass given by the Casimir term and that the thermodynamic volume is always positive if the regularity condition is applied.

 The outline of our paper is as follows: in the next section we introduce the EH Solitons in odd dimensions. We briefly discuss general considerations of their thermodynamics  in section 3. In section 4, we make use of the first law and Smarr relation to
 compute the mass and thermodynamic volume of these solutions inside and outside the cosmological horizon of dS space . We show that explicit expressions for the two parameters can be found in general odd dimensions.  We briefly summarize our results in a concluding section.


\section{EHdS solitons}

EHdS solitons \cite{Clarkson:2005qx, Clarkson:2006zk} in general odd $(d+1)$ dimensions 
are exact solutions to the Einstein equations with $\Lambda > 0$, and have metrics of the form
\begin{equation}
ds^{2} = -g(r)dt^{2}+\left( \frac{2r}{d}\right) ^{2}f(r)\left[ d\psi
+\sum_{i=1}^{k}\cos (\theta _{i})d\phi _{i}\right] ^{2}   
+\frac{dr^{2}}{g(r)f(r)}+\frac{r^{2}}{d}\sum_{i=1}^{k}d\Sigma _{2(i)}^{2}
\label{EH d-dim}
\end{equation}
in $d=2k+2$ dimensions, where the metric functions  are given by
\begin{equation}
g(r)=1 - \frac{r^{2}}{\ell ^{2}}~~~,~~~~~f(r)=1-\left( \frac{a}{r}\right) ^{d}
\label{d-dimmetricfns}
\end{equation}%
with 
\begin{equation}
d\Sigma _{2(i)}^{2}=d\theta _{i}^{2}+\sin ^{2}(\theta _{i})d\phi _{i}^{2}
\label{dOmegasq}
\end{equation}%
and 
\begin{equation}
\Lambda = + \frac{d(d-1)}{2\ell ^{2}} 
\label{lambdanorm}
\end{equation}
parametrizing the positive cosmological constant.

The radial coordinate $r \geq a$; for $r<a$ the metric changes signature, indicative of its solitonic character. There is a cosmological horizon at $r=\ell$. Constant $(t,r)$ sections consist of the fibration of a circle over a product of $k$ 2-spheres.
Generalizations to Gauss-Bonnet gravity  \cite{Wong:2011aa}  and to spacetimes with more general base spaces \cite{Mann:2005ra} exist
but we shall not consider these solutions here.

For $\ell \rightarrow \infty $, the metric \eqref{EH d-dim} becomes
\begin{equation}
ds^{2}=\left( \frac{2r}{d}\right) ^{2}\left( 1-\left( \frac{a}{r}\right)
^{d}\right) \left[ d\psi +\sum_{i=1}^{k}\cos (\theta _{i})d\phi _{i}\right]
^{2}+\frac{dr^{2}}{1-\left( \frac{a}{r}\right) ^{d}}+\frac{r^{2}}{d}
\sum_{i=1}^{k}d\Sigma _{2(i)}^{2}  \label{ddimEH}
\end{equation}
for a constant $t=$ hypersurface. This class of metrics can be regarded as $d$-dimensional generalizations of the original \cite{Eguchi:1978xp} $d=4$ Eguchi-Hanson metric.

In general, the metric \eqref{EH d-dim} will not be regular unless some conditions are imposed to eliminate the singularities. Noting that a constant $(t,r)$ section has the form
\begin{equation}
ds^{2}=F(r)\left[ d\psi +\sum_{i=1}^{k}\cos (\theta _{i})d\phi _{i}\right]
^{2}+\frac{dr^{2}}{G(r)}
\end{equation}%
where $F(r) =\left( \frac{2r}{d}\right) ^{2} f(r)$ and $G(r) = f(r) g(r)$,  regularity requires the absence of conical singularities. This implies that the periodicity of $\psi$ at infinity must be an integer multiple of its periodicity as $r\rightarrow a$. Consequently
\begin{equation}
\left.\frac{4\pi }{\sqrt{\left| F^{\prime }G^{\prime }\right| }}\right\vert_{r=a}=\frac{%
4\pi }{p}  
\label{regularity}
\end{equation}%
where $p$ is an integer. Note that $r=r_{+}$ is the simultaneous root of $F$ and
$G$ and that $F_{+}^{\prime }G_{+}^{\prime }=4\left(
1 - \frac{a^{2}}{\ell ^{2}}\right) $.

The implications of the regularity condition vary  depending on the following three cases: \ $a^{2}<\ell ^{2}$, $a^{2}>\ell ^{2}$ and $a^{2}=\ell ^{2}$.  
If $a^{2}<\ell ^{2}$,  the regularity condition yields $p=1$ and thus $a^{2}=\frac{3}{4}\ell ^{2}$.  
If $a^{2}>\ell ^{2}$, when $\ell<r<a$, the metric has closed timelike curves.  If  $a=\ell $ the metric is not static for $r>a$.  We shall not consider these latter two cases in this paper.

In the sequel we shall investigate the thermodynamic behavior of the metric \eqref{EH d-dim} for general values of $a< \ell$, imposing the regularity condition  $a^{2}=\frac{3}{4}\ell ^{2}$ at the end of the calculation.  This will allow us to explore the thermodynamics of a cosmological horizon in thermodynamic equilibrium under rather general conditions without any complicating features due to the presence of a black hole.


\section{Soliton Thermodynamics}

Since the Killing vector $\partial/\partial t$ is not everywhere timelike, we cannot compute the mass 
$M$ of the soliton unambiguously.
As a consequence we cannot directly compute the thermodynamic volume $V =\frac{\partial M}{\partial P}$ without additional
assumptions.   We shall assume the validity of the first law \eqref{firstgenD} and the Smarr relation \eqref{SmarrgenD} to compute their mass and the volume.  This approach is analogous to that taken for asymptotically Lifshitz black holes \cite{Brenna:2015pqa}, for which computation of
the mass is also fraught with ambiguity in certain cases. We shall then relate our computation of the mass to that obtained in other procedures.

 In $(d+1)$ spacetime dimensions,  the entropy of the EHdS soliton follows from the area law
 \begin{equation}\label{entrop}
S = K \ell^{(d-1)} \sqrt{1-{\left(\frac{a}{\ell}\right)}^d}
 \end{equation}
with standard arguments implying  the temperature at the cosmological horizon is
 \begin{equation}\label{temp}
T= \frac{\sqrt{1-{\left(\frac{a}{\ell}\right)}^d}}{2 \pi \ell}
 \end{equation}
where $K=  \frac{1}{2p}\left(\frac{4\pi} {d}\right)^{\frac{d}{2}}$ is one-quarter of the area of the cosmological horizon when $a=0$.
   
Before proceeding we note that    \eqref{firstgenD} and  \eqref{SmarrgenD} determine the mass and volume for any  solution to the field equations only up to an additive term that depends on $\ell$. Using \eqref{press} and 
\eqref{lambdanorm} it is straightforward to compute this contribution
 \begin{equation}\label{MVag}
M_\Delta = \alpha_{d} \ell^{d-2}  \qquad  
V_\Delta = \frac{|\Lambda|}{\Lambda} \frac{8\pi (d-2)}{d(d-1)}  \alpha_{d}  \ell^d
 \end{equation}
 for both the AdS and dS cases, where $\alpha_d$ is an arbitrary constant.  Note that the respective  contributions to the mass and volume have opposite signs in the AdS case but the same sign in the de Sitter case.
 
These additional terms depend only on $\ell$, suggesting they be considered as  Casimir contributions to the mass and volume.  However this interpretation is fraught with problems in the AdS case for several reasons.  First, they are present in any spacetime dimension, whereas Casimir contributions to the mass occur only for odd spacetime dimensions (even $d$), and so this interpretation is inapplicable for odd $d$.    Second, they alternate in sign:  for $d=4,6,8$ 
it has been shown that  $\alpha_d = 3\pi/32, -5\pi^2/128, 35\pi^3/3072$ respectively \cite{Das:2000cu}.  
This necessarily yields a negative contribution to the volume for $d=4n$ where $n$ is an integer, 
and these contributions can make the overall volume of a sufficiently small Schwarzschild Anti de Sitter black hole negative. Finally, there is no sensible $\Lambda\to 0$ (or $\ell\to\infty$) limit of these contributions unless $\alpha_d=0$.  For these reasons the constant $\alpha_d$ is generally set to zero for asymptotically AdS solutions.

However in the de Sitter case,  \eqref{firstgenD} and  \eqref{SmarrgenD} imply that $M_\Delta$ and $V_\Delta$ have the same sign, and it is not clear
that such contributions  should be set to zero.  For the soliton  solutions we are considering, $d$ is always even and so it is reasonable to expect a Casimir contribution to the cosmological volume.  
Indeed we shall see that a variety of interpretations for this additional term
exist, and we shall explore a number of distinct possibilities.

  We solve the first law and Smarr relation for the conserved mass and the cosmological volume of EHdS solitons in both cases: inside and outside the cosmological horizon.  On dimensional grounds we expand the mass $M$ and the cosmological volume $V_c$ in powers of $a$ and $\ell$
 \begin{equation}\label{mseries}
M= \sum_{k=0}^{\frac{d}{2}} m_k a^{d-2k} \ell^{2k-2}
\end{equation}  
and
\begin{equation}\label{vseries}
 V = \sum_{k=0}^{\frac{d}{2}} v_k a^{d-2k} \ell^{2k} 
\end{equation}
which are the most general expansions admitting  a solution that satisfies both \eqref{firstgenD} and \eqref{SmarrgenD}.  In fact it is more than we need -- noting that
 the $\ell$-dependent term in the mass is divergent in the limit $\Lambda\to 0$  suggests that we should consider 
excluding it.  However since there is no soliton in this limit, we have retained this term.  We shall investigate the implications of identifying it with the Casimir energy \cite{Ghezelbash:2001vs}
in the dS/CFT correspondence conjecture \cite{Strominger:2001pn}.

\subsection{Inside the cosmological horizon}

The Smarr  relation (\ref{SmarrgenD}) and the first law (\ref{firstgenD}) are valid for
a black hole horizon.  For a cosmological horizon these equations are modified to read \cite{Dolan:2013ft}   
\begin{equation}\label{FLin}
\delta M_{in} + T{dS}  - V_{in} \delta P = 0
\end{equation}
\begin{equation}\label{Smarrin}
(d-2) M_{in} + (d-1) TS  + 2 V_{in} P = 0
\end{equation}
from the perspective of an observer in a region where $\partial/\partial t$ is timelike, and where the plus signs in the second terms of these equations arises because the surface gravity of the de Sitter horizon is negative, while the corresponding  temperature $T  > 0$ since it is proportional to the magnitude of the surface gravity.

Using \eqref{entrop} and \eqref{temp}, we find that most terms in both \eqref{mseries} and \eqref{vseries} vanish and obtain
 \begin{equation}\label{MVin}
M_{in}=  \frac{K a^d}{4 \pi \ell^2} + m_{\frac{d}{2}} \ell^{d-2}
 \qquad
 V_{in} =  \frac{-2 K}{d-1} a^d + \left(\frac{8(d-2)}{d(d-1)} \pi m_{\frac{d}{2}} + \frac{4 K}{d} \right) \ell^d
\end{equation}
for the mass and volume respectively, where we have relabeled $\alpha_d \to m_{\frac{d}{2}}$.

If we impose the requirement that  the mass remain finite as $\Lambda\to 0$, then $m_{\frac{d}{2}} =0$ and   
\begin{eqnarray}\label{Min-2}
M_{in} &=&  \frac{\left(\frac{4\pi}{d}\right)^\frac{d}{2} a^d}{8 p \pi  \ell^2} \to M_{sol} =  \frac{1}{8 \pi} \left(\frac{ 3\pi}{d}\right)^{d/2}\ell^{d-2} 
\\
 V_{in} &=&   \frac{2 \left( \frac{4\pi}{d}\right)^\frac{d}{2} }{p d} \ell^d \left( 1- \frac{d }{2(d-1)} \frac{a^d}{\ell^d} \right) 
\to V_{sol} =  \frac{2 \ell^d \left( \frac{4\pi}{d}\right)^\frac{d}{2} }{d}   \left( 1- \frac{d }{2(d-1)} \left(\frac{ 3}{4}\right)^{\frac{d}{2}} \right) \quad
\end{eqnarray}
\label{Vin-2}
where the latter relations follow from imposing the regularity condition \eqref{regularity}.

Suspending the regularity condition, we note that for any value of $a<\ell$ the mass $M_{in}$ and cosmological volume $V_{in}$ in \eqref{MVin} are both positive for vanishing $ m_{\frac{d}{2}}$.  However the volume $V_{in}$ does not vanish in the $\Lambda \to 0$ limit, and so one might consider a variety of choices of $m_{\frac{d}{2}}$ that will yield various desired outcomes.  These we depict in tables 1 and 2. In table 1 we have not imposed the regularity condition  \eqref{regularity}, and indicate
the choices of $m_{\frac{d}{2}}$ such that
the rows correspond to finite mass as $\ell\to \infty$ (case 1), vanishing mass (case 2), vanishing volume (case 3), and requiring the volume to depend only on the soliton parameter $a$ (case 4).  In table 2 we impose the regularity condition, the rows in this table corresponding to those in table 1 for the respective choices of $m_{\frac{d}{2}}$.
\begin{figure}[htp]
\centering
\includegraphics[width=0.32\textwidth]{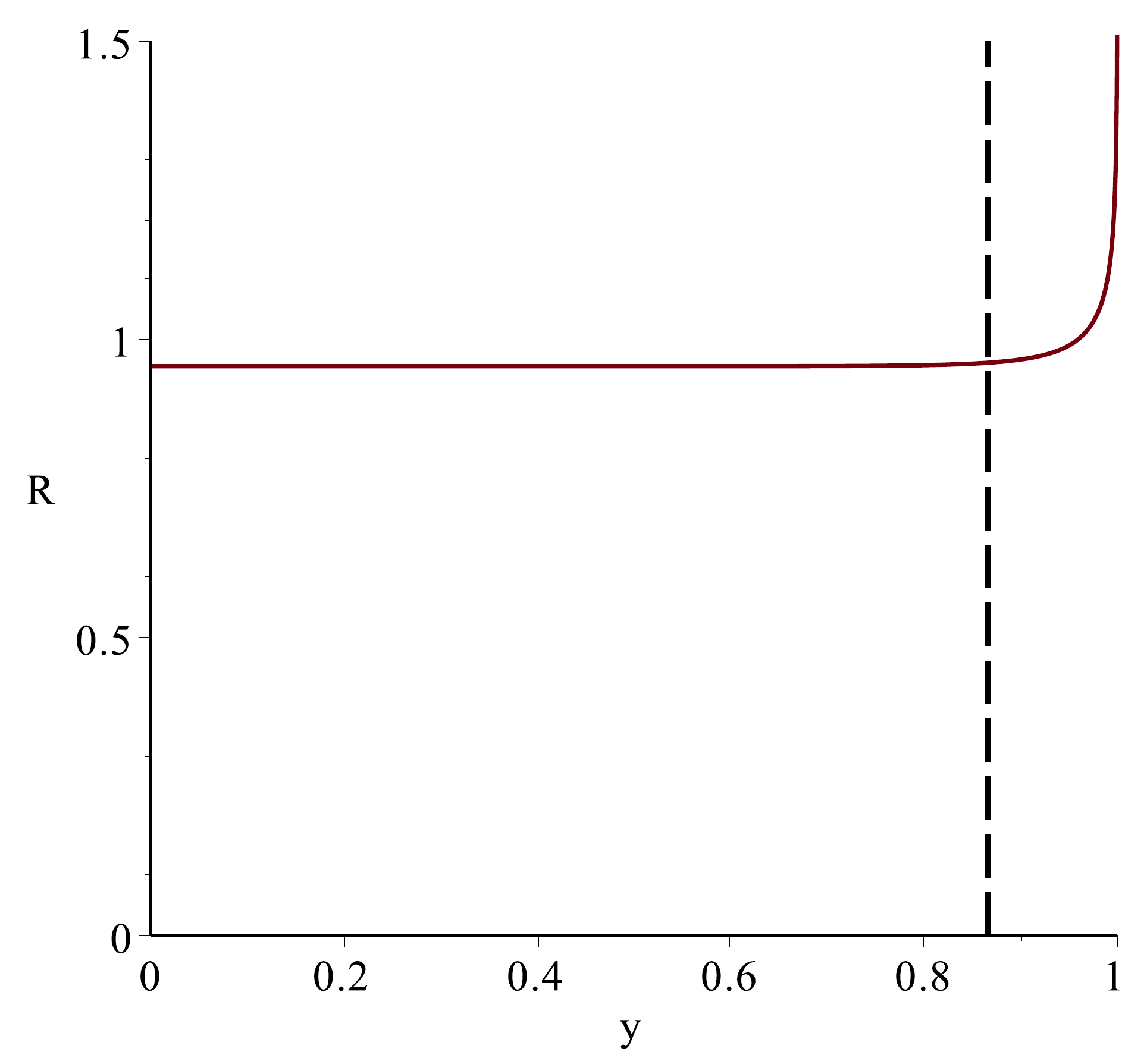}
\includegraphics[width=0.32\textwidth]{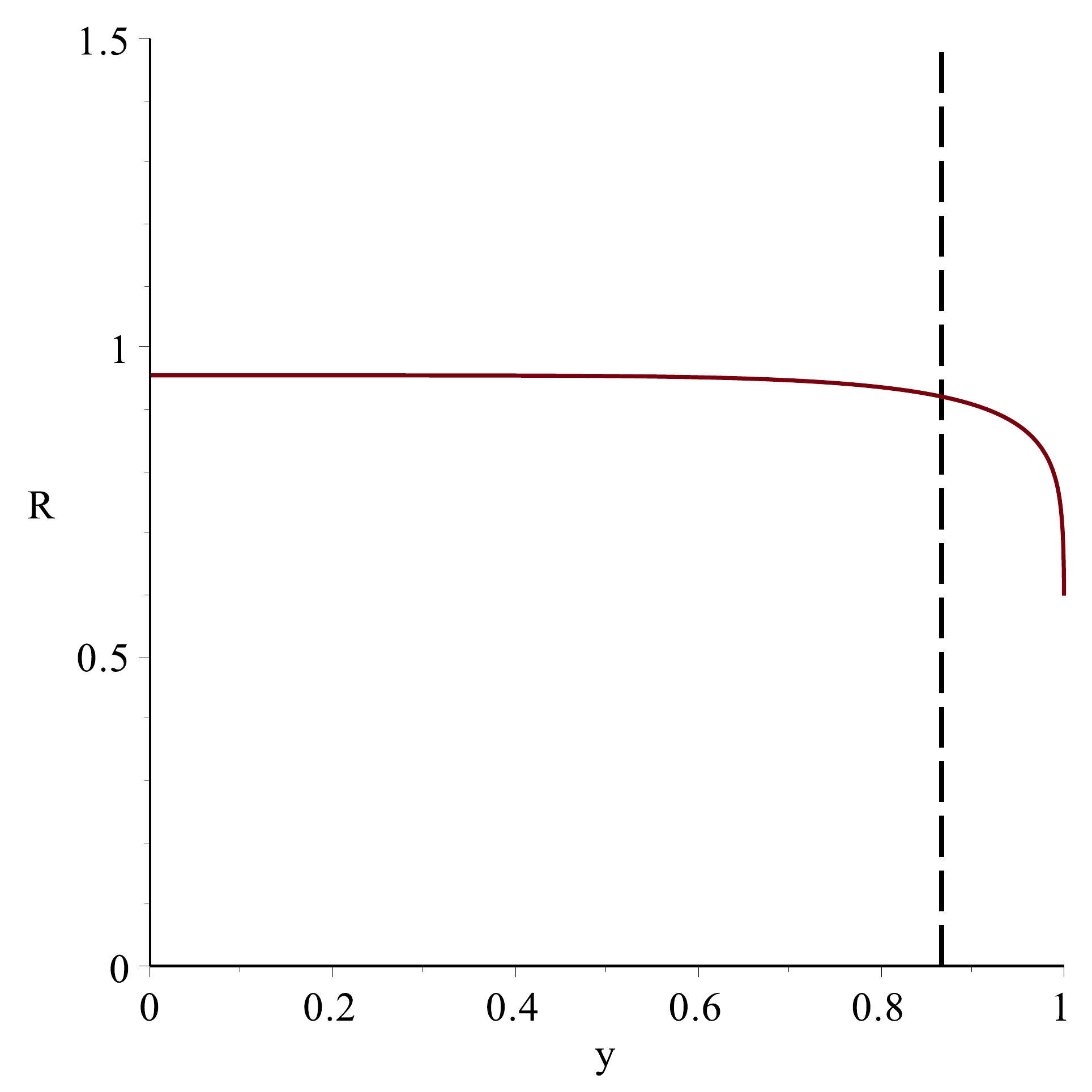}
\includegraphics[width=0.32\textwidth]{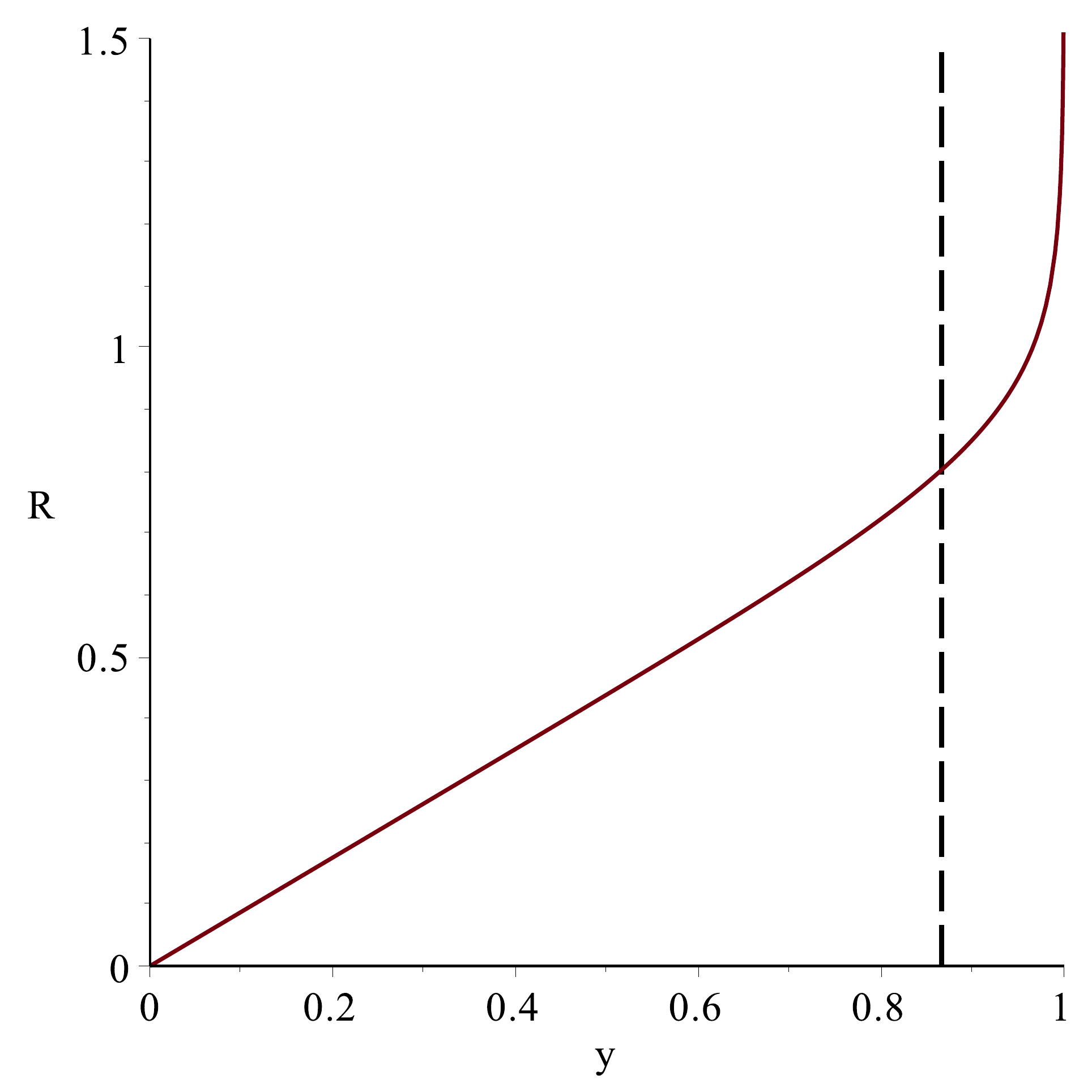}
\caption{Plots of the isoperimetric ratio $\mathcal{R}$ (for a representative dimension $d=6$) as a function of the parameter $y=\frac{a}{\ell}$. The plots, from left to right, correspond respectively to  case 1,  case 2, and  case 4 discussed in Table \ref{tab:first-table}.  The dashed line corresponds to the regularity condition
$y=\sqrt{3/4}$ . }
\label{lambda-effects}
\end{figure}

\begin{figure}[htp]
\centering
\includegraphics[scale=0.38]{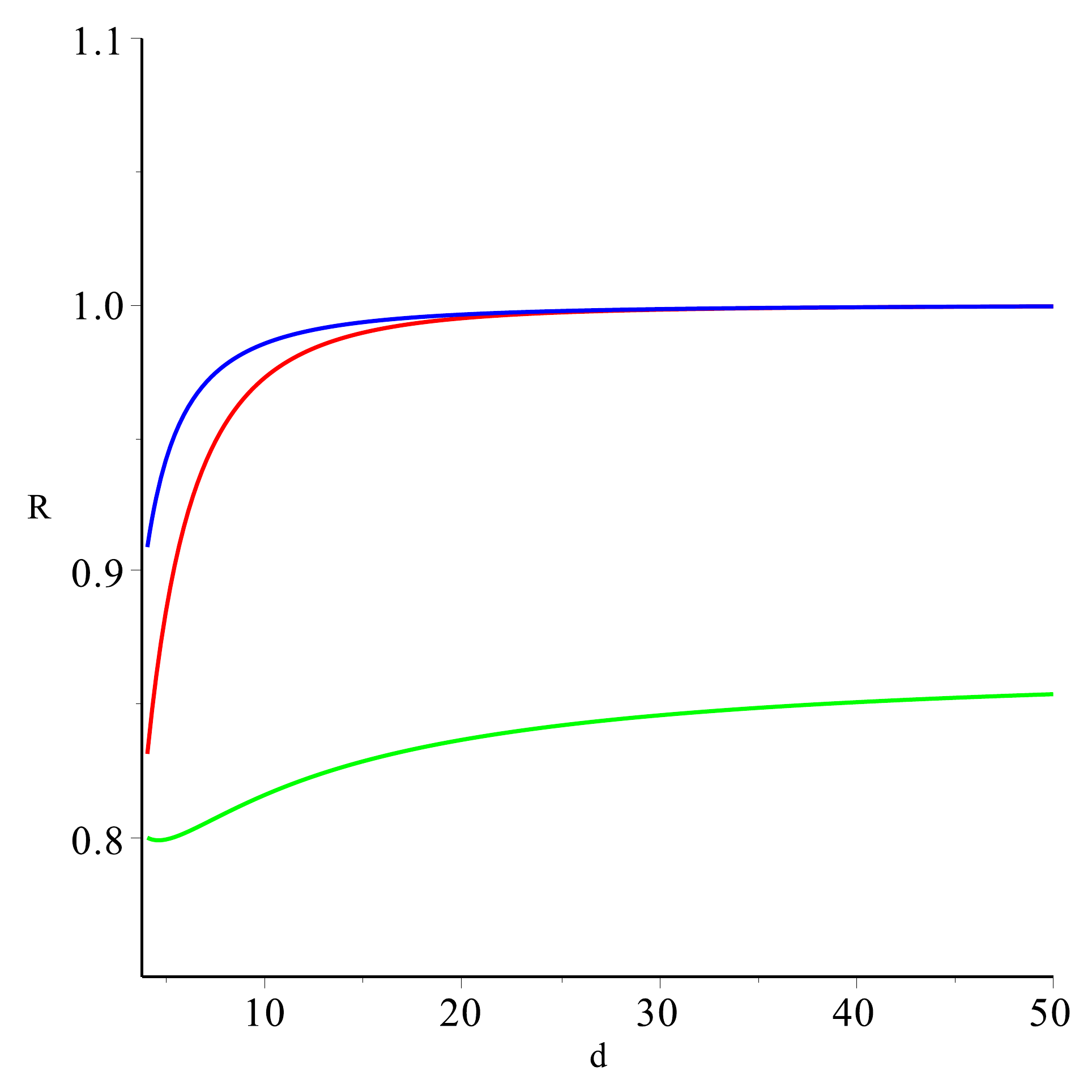}
\caption{A plot of the isoperimetric factor $\mathcal{R}$ for the cases 1, 2 and 4 of the cosmological volume as a function of  the spatial dimension $d$ when the regularity condition is imposed (see Table \ref{tab:second-table}).  The first case is (blue), the second is in (red) and the fourth is in (green).
}
\label{fig:R_general_d}
\end{figure}

Since the soliton is not a black hole, the premises of the Reverse Isoperimetric Inequality Conjecture \cite{Cvetic:2010jb} do not apply.  However the parameter $\mathcal{R}$ provides a useful measure of the relationship between volume and entropy, and so
we indicate in each table a computation of the isoperimetric ratio $\mathcal{R}$ in \eqref{eq:ipe-ratio} .  We find that $\mathcal{R}$ is always less than unity if the regularity condition is imposed, as indicated in  table 2 and illustrated in figures \ref{lambda-effects} and \ref{fig:R_general_d}. 
If the regularity condition is not imposed then for a small range of values of $a > \sqrt{3/4}\ell$ we can obtain $\mathcal{R} > 1$, as shown in figure \ref{lambda-effects}.

\hvFloat[floatPos=t!, rotAngle=0,capWidth=w]%
{table}%
{
\begin{adjustbox}{max width=\textwidth}
\begin{tabular}{|| l| c| c| c| c||} 
  \hline
  & $m_{\frac{d}{2}}$ & $M_{in}$ & $V_{in}$ & $\mathcal{R}\left(y=\frac{a}{\ell}\right)$ \\ [1.5ex] 
 \hline\hline
 Case 1: Finite ${M}_{in}$ as $\ell \to \infty$ & 0 & $\frac{K a^d}{4 \pi \ell^2}$ & $\frac{-2K}{d-1}a^d + \frac{4K}{d}\ell^d$ & $\left( 4 - \frac{2d}{d-1} y^d \right)^\frac{1}{d} \left( \frac{1}{4\sqrt{1-y^d}}\right)^\frac{1}{d-1} $ \\[1.5ex]
 \hline
 Case 2: Vanishing Mass & $\frac{-K}{4\pi} \left( \frac{a}{\ell} \right)^d$ & 0 & $\frac{4K}{d} \left( \ell^d - a^d\right)$ & $\left( 4 - 4 y^d \right)^\frac{1}{d} \left( \frac{1}{4\sqrt{1-y^d}}\right)^\frac{1}{d-1} $ \\ [1.5ex]
 \hline
Case 3: Vanishing Volume  & $\frac{K}{2\pi}\frac{d-1}{d-2} + \frac{K d}{4\pi(d-2)} \left(\frac{a}{\ell}\right)^d$ &  $\frac{K}{2\pi} \frac{(d-1)}{(d-2)} \left(  \frac{a^d}{\ell^2} - \ell^{d-2} \right)$ & 0  & 0 \\ [1.5ex]
 \hline
Case 4: $V_{in}$ as function of $a$ & $\frac{-K}{2\pi}\frac{d-1}{d-2}$ &  $\frac{-K}{4\pi} \left(  \frac{2(d-1)}{(d-2)} \ell^{d-2} - \frac{a^d}{\ell^2} \right)$ & $ \frac{-2 K}{d-1} a^d$  & $ y \left( \frac{2d}{d-1} \right)^\frac{1}{d} \left( \frac{1}{4\sqrt{1-y^d}}\right)^\frac{1}{d-1} $ \\ [1.5ex]
 \hline
\end{tabular}
\end{adjustbox}
}%
{General results }%
{tab:first-table}

\hvFloat[floatPos=b!,rotAngle=0,capWidth=w]%
{table}%
{ 
\begin{adjustbox}{max width=\textwidth}
\begin{tabular}{|| l| c| c| c| c||} 
  \hline
  & $m_{\frac{d}{2}}$ & $M_{sol}$ & $V_{sol}$ & $\mathcal{R}\left(d = 6\right)$ \\ [1.5ex] 
 \hline\hline
Case 1: Finite ${M}_{in}$ as $\ell \to \infty$ & 0 & $\frac{1}{8\pi} \left( \frac{3\pi}{d} \right)^\frac{d}{2} \ell^{d-2}$ & $\left[ - \frac{1}{d-1} \left( \frac{3\pi}{d} \right)^\frac{d}{2} + \frac{1}{8 d} \left(\frac{4 \pi}{d}\right)^{\frac{d}{2}}\right] \ell^d$ & $ 0.96073$ \\[1.5ex]
 \hline
 Case 2: Vanishing Mass & $- \frac{1}{8 \pi}\left( \frac{3\pi}{d} \right)^\frac{d}{2}$ & 0 & $- \frac{1}{8 d}\left( \frac{4\pi}{d} \right)^\frac{d}{2} \left( 1 - \left(\frac{3}{4}\right)^\frac{d}{2} \right) \ell^d$ & $0.92058 $ \\ [1.5ex]
 \hline
  Case 3: Vanishing Volume  & $\frac{1}{4 \pi \left(d-2\right)}\left( \frac{4\pi}{d} \right)^\frac{d}{2}\left[d-1+\frac{d}{2}\left(\frac{3}{4}\right)^\frac{d}{2}\right]$ &  $\frac{1}{4 \pi} \frac{d-1}{d-2}\left( \frac{4\pi}{d} \right)^\frac{d}{2}  \left[ \left(\frac{3}{4}\right)^\frac{d}{2} - 1 \right] \ell^{d-2}$ & 0  & 0 \\ [1.5ex]
 \hline
Case 4: $V_{in}$ as fonction of $a$ & $- \frac{1}{4 \pi} \frac{d-1}{d-2}\left( \frac{4\pi}{d} \right)^\frac{d}{2}$ &  $-\frac{1}{8\pi} \left( \frac{4\pi}{d} \right)^\frac{d}{2} \left[ 2 \frac{d-1}{d-2} - \left(\frac{3}{4}\right)^\frac{d}{2} \right] \ell^{d-2}$  & $ - \frac{1}{d-1} \left( \frac{3\pi}{d} \right)^\frac{d}{2}  \ell^d$ & $0.80220$\\ [1.5ex]
 \hline
\end{tabular}
\end{adjustbox}
}%
{Results with the regularity condition $y^2=\frac{3}{4}$  imposed
}%
{tab:second-table}

\subsection{Outside the cosmological horizon}

In this section we consider the thermodynamics of EH-dS solitons outside of the cosmological horizon,
where $r>\ell$.  This problem (without taking thermodynamic volume into account) has been previously considered in the context of the proposed dS/CFT correspondence, which entails computing quantities at past/future infinity. The timelike Killing vector $\xi=\partial /\partial t$ becomes
spacelike and its associated conserved charge   can be
calculated using the relationship
\begin{equation}
\mathfrak{M} = \mathfrak{Q}_{\xi }=\oint_{\Sigma }d^{d-1}S^{a}\xi ^{b}T_{ab}^{\text{eff}}
\label{Qconsgen}
\end{equation}
where  $T_{ab}^{\text{eff}}$ is  stress-energy on the boundary $\Sigma$ of the manifold, determined from varying the
Einstein-dS action with counter-terms.  Full expansions for this quantity have been previously computed \cite{Ghezelbash:2001vs}.  Using \eqref{Qconsgen}
the maximal mass
conjecture ---   \textit{any asymptotically dS spacetime with mass greater than dS has a cosmological singularity} --- was proposed \cite{Balasubramanian:2001nb}. 
A straightforward evaluation of  \eqref{Qconsgen} at future infinity
for the Schwarzschild de Sitter solution 
whose metric functions in the ansatz \eqref{EH d-dim} are $g(r) = 1-r^2/\ell^2 - 2m/r^2$, $f(r)=1$ for
$d=4$ yields $\mathfrak{M} = - m$  \cite{Clarkson:2003kt}, which is sign-reversed from the  quantity employed  in \eqref{FLin} and \eqref{Smarrin}.   This is a general property of  computing
the mass outside of a cosmological horizon using \eqref{Qconsgen} \cite{Clarkson:2003kt}.

Hence in order to apply our approach, we make use of  the Smarr  relation (\ref{Smarrin}) and the first law (\ref{FLin})  with $M\to -M$ to solve for the outer mass and the cosmological volume.  The net effect of this is to recover the relations  (\ref{SmarrgenD}) and  (\ref{firstgenD}) 
but with $V\to -V$:
\begin{equation}\label{FLout}
\delta M_{out} -T dS + V_{out} \delta P = 0
\end{equation}
\begin{equation}\label{Smarrout}
(d-2) M_{out} - (d-1) TS - 2 V_{out} P = 0
\end{equation}
and we shall solve these for $M_{out}$ and $V_{out}$
using \eqref{entrop} and \eqref{temp}. 

The calculation is very similar to the inside case.  We obtain
 \begin{equation}\label{M_out}
M_{out}=  - \frac{K a^d}{4 \pi \ell^2} + m_{\frac{d}{2}} \ell^{d-2}
\end{equation}  
and
\begin{equation}\label{V_out}
 V_{out} = -\frac{2 K}{d-1} a^d - \left(\frac{8(d-2)}{d(d-1)} \pi m_{\frac{d}{2}} - \frac{4 K}{d} \right) \ell^d
\end{equation}
as the general solutions to (\ref{SmarrgenD}) and (\ref{firstgenD}). We see that in general the
contribution proportional to $m_{\frac{d}{2}}$ is now of opposite sign for the mass and volume.

We can compare our results in $d=4$ to a direct computation of the mass.  
The EHdS metric reads in this case
 \beq
ds^2 = - g(r) dt^2 + \frac{r^2}{p^2} f(r) \left[ d\Psi + \frac{p}{2} \cos\theta d\Phi \right] ^2 + \frac{1}{f(r) g(r)} dr^2 + \frac{r^2}{4} d{\Omega_2}^2
\eeq
\label{EH 5-dim}
where  the counter-term method \cite{Ghezelbash:2001vs} yields  
\begin{equation}\label{mass-count}
\mathfrak{M} = \frac{\pi \left( 3 \ell^4 - 4 a^4  \right)}{32 p  \ell^2}
\end{equation}
for  the conserved mass using  \eqref{Qconsgen} \cite{Clarkson:2005qx}.  The action can likewise be directly computed in this approach and is 
\begin{equation}
I = \frac{\beta \pi \left( 4 a^4 - 5 \ell^4  \right)}{32p \ell^2}
\end{equation}
yielding  the entropy 
\begin{equation}\label{ent-count}
S = \frac{\beta \pi \left( \ell^4  -a^4 \right)}{4p \ell^2}
\end{equation}
via the Gibbs-Duhem relation $S = \beta M - I $, where  
\begin{equation}\label{beta}
\beta = \frac{2 \pi \ell^3}{\sqrt{\ell^4 - a^4}}
\end{equation}
is the period of the Euclidean time $\tau$ that ensures regularity in the $(\tau,r)$ section of
the Euclidean solution. Note that the regularity condition has not been applied.

Clearly for $d=4$, $\beta= 1/T$ from \eqref{temp}, and the entropy \eqref{ent-count} agrees with
\eqref{entrop}.  Requiring that these quantities along with the mass \eqref{mass-count} satisfy both
the first law \eqref{firstgenD} and Smarr relation \eqref{SmarrgenD} yields
\begin{equation}\label{vol-out}
V_{out} = \frac{\pi^2}{24p} \left(  9 \ell^4 -8 a^4 \right)
\end{equation}
for the thermodynamic volume of the EHdS soliton, where from \eqref{press}
\begin{equation}
P = -\frac{\Lambda}{8 \pi G} = -\frac{3}{4 \pi \ell^2}
\end{equation}
is the pressure.

\hvFloat[floatPos=t!, rotAngle=0,capWidth=w]%
{table}%
{
\begin{tabular}{|| c| c| c| c||} 
  \hline
Dimension & $m_{\frac{d}{2}}$ & $M_{out}$ & $V_{out}$  \\ [1.5ex] 
 \hline\hline
 $d=4$ & $\frac{3 \pi}{32}$ & $\frac{3 \pi}{128} \ell^2$ & $ \frac{3 \pi^2}{16} \ell^4$  \\[1.5ex]
 \hline
 $d=6$ & $\frac{5 \pi^2}{128}$ & $\frac{3 \pi^2}{128} \ell^2$ & $ \frac{13 \pi^3}{405} \ell^6$  \\ [1.5ex]
 \hline
  $d=8$ & $\frac{35 \pi^3}{3072}$ & $\frac{877 \pi^3}{98304} \ell^2$ & $ \frac{87 \pi^4}{28672} \ell^8$  \\ [1.5ex]
 \hline
\end{tabular}
}%
{Mass and Volume computed outside the cosmological horizon.  The constant $m_{\frac{d}{2}}$ is chosen to yield the mass of de Sitter spacetime when $a=0$. }%
{tab:third-table}


By imposing the regularity condition $p=1$, we obtain
\begin{equation}\label{MVout}
M_{out} =  
 \frac{3\pi  \ell^2}{128} \qquad 
V_{out} =  \frac{3\pi^2  \ell^2}{16}
\end{equation}
which is in agreement with both the mass \eqref{M_out} and the cosmological volume \eqref{V_out} if we set  $m_{2} = \frac{3 \pi}{32}$ (and noting that $K=\pi^2/2p$ for $d=4$).
We note from \eqref{MVout} that  $M_{out} < M_{dS} =  \frac{3\pi  \ell^2}{32} $, in accord with the maximal mass conjecture \cite{Balasubramanian:2001nb,Clarkson:2003kt,Clarkson:2004yp}, and  that this yields a positive thermodynamic volume.  If we choose
$M_{out} = M_{dS}$, then $m_{2} = \frac{21 \pi}{128}$, and the volume is
still positive.  A negative  volume  requires $m_{2} > \frac{15 \pi}{64}$, which then yields
$M_{out} =  \frac{21 \pi \ell^2}{128} > M_{dS}$, in violation of the conjecture.

Similar arguments for $d>4$ can be made. It is always possible to choose the constant
$m_{\frac{d}{2}}$ to yield  $M_{out} =  M_{dS}$ when $a=0$, and it is clear from  \eqref{MVout}
that the maximal mass conjecture will necessarily be satisfied.  As shown in table \ref{tab:third-table}, we find
for all values that have been calculated for $M_{dS}$ \cite{Ghezelbash:2001vs}
that the volume $V_{out} > 0$.  We expect
that this is a general feature for any (odd) dimension.


\section{Conclusion}

 We have found general expressions for the thermodynamic volume inside and outside the cosmological horizon for EH solitons in any odd dimension.  These quantities are calculable and well-defined regardless of whether or not the regularity condition for the soliton is satisfied.  They illustrate that cosmological volume is a well-defined concept, and that cosmological horizons  indeed have meaningful thermodynamic properties. 

For observers within the cosmological horizon, the mass and volume can be defined using the first law and Smarr relations.  We  have shown that for this case that the reverse isoperimetric inequality \cite{Cvetic:2010jb} is not satisfied  for general values of the soliton parameter $a$ (including the value
satisfying the regularity condition), though it is satisfied for a narrow range of values of this parameter.
This situation stands in contrast to that for the class of Kerr de Sitter spacetimes, for which 
$\mathcal{R}\ge 1$ holds for cosmological horizons \cite{Dolan:2013ft}. That $\mathcal{R}$ is less than unity even when the soliton regularity condition is satisfied hints at a   relationship between the degrees of freedom of cosmological horizons and their entropy that is distinct from that of black holes.

For the outer case we exploited the definition \eqref{Qconsgen} of conserved mass to 
obtain the unique result \eqref{vol-out} for the cosmological volume in 5 dimensions (or alternatively
\eqref{MVout} when the regularity condition holds).  The mass $M_{out}$ satisfies the maximal mass conjecture and the volume is positive.  By $M_{out}$ to yield the mass 
 \eqref{Qconsgen} for de Sitter space when $a=0$, we find that the associated cosmological volume 
 is always positive in all dimensions for which \eqref{Qconsgen} has been computed.  We expect this is a general feature for all spacetimes satisifying the maximal mass conjecture.
  
The thermodynamics of these objects remains to be further explored.  The equation of state for
the (non-regular) soliton will, from \eqref{temp}, be a highly non-linear relationship between the pressure, volume, and temperature, and  whether or not any interesting phase behaviour can result remains to be determined.  Generalizations of these solutions to Lovelock gravity exist \cite{Wong:2011aa}, and it is quite possible that these objects may also exhibit interesting thermodynamic behaviour.

\section{Acknowledgements}
We would like to thank \textbf{\textit{ la Mission Universitaire Tunisienne en Amerique du Nord}} (Mbarek) and \textbf{\textit{ the Natural Sciences and Engineering Research Council of Canada}} for the financial support.

\bibliography{LBIB2}
\bibliographystyle{JHEP}
\end{document}